\documentclass{SciPost}

\hypersetup{
    colorlinks,
    linkcolor={red!50!black},
    citecolor={blue!50!black},
    urlcolor={blue!80!black}
}

\usepackage[bitstream-charter]{mathdesign}
\usepackage{amsmath}
\usepackage{subcaption}

\DeclareSymbolFont{usualmathcal}{OMS}{cmsy}{m}{n}
\DeclareSymbolFontAlphabet{\mathcal}{usualmathcal}

\fancypagestyle{SPstyle}{
\fancyhf{}
\lhead{\colorbox{scipostblue}{\bf \color{white} ~SciPost Physics }}
\rhead{{\bf \color{scipostdeepblue} ~Submission }}

\fancyfoot[C]{\textbf{\thepage}}
}

\begin{document}

\pagestyle{SPstyle}

\begin{center}{\Large \textbf{\color{scipostdeepblue}{
Spectroscopic Signatures of a Liouvillian Exceptional Spectral Phase in a Collective Spin\\
}}}\end{center}

\begin{center}\textbf{
Rafael A. Molina
}\end{center}

\begin{center}
{\bf 1} Instituto de Estructura de la Materia (IEM-CSIC), E-28006, Madrid, Spain
\\[\baselineskip]
\href{mailto:rafael.molina@csic.es}{\small rafael.molina@csic.es}
\end{center}

\section*{\color{scipostdeepblue}{Abstract}}
\textbf{\boldmath{%
Non-Hermitian degeneracies of Lindblad generators (Liouvillian exceptional points) can induce
non-exponential relaxation and higher-order poles in dynamical response functions. A collective spin coupled to a polarized Markovian bath exhibits an \emph{exceptional spectral phase}
in which defective Liouvillian modes imprint super-Lorentzian features in frequency-resolved spectra.
We compute the emission spectrum via the Liouvillian resolvent, identify symmetry-sector selection rules,
and demonstrate that exceptional-point signatures are strongly state-dependent: they are suppressed in
steady-state fluorescence yet become unambiguous for generic (infinite-temperature or random) initial
states. Our results provide an experimentally accessible spectroscopic diagnostic of many-body
Liouvillian exceptional phases and clarify when steady-state emission can (and cannot) reveal them.
}}

\vspace{\baselineskip}

\noindent\textcolor{white!90!black}{%
\fbox{\parbox{0.975\linewidth}{%
\textcolor{white!40!black}{\begin{tabular}{lr}%
  \begin{minipage}{0.6\textwidth}%
    {\small Copyright attribution to authors. \newline
    This work is a submission to SciPost Physics. \newline
    License information to appear upon publication. \newline
    Publication information to appear upon publication.}
  \end{minipage} & \begin{minipage}{0.4\textwidth}
    {\small Received Date \newline Accepted Date \newline Published Date}%
  \end{minipage}
\end{tabular}}
}}
}


\vspace{10pt}
\noindent\rule{\textwidth}{1pt}
\tableofcontents
\noindent\rule{\textwidth}{1pt}
\vspace{10pt}


\section{Introduction}
\label{sec:intro}
Exceptional points (EPs)---non-Hermitian degeneracies at which both eigenvalues and eigenvectors coalesce---have emerged as a unifying concept behind a broad range of critical phenomena in open and effective non-Hermitian systems. Formally, EPs are branch-point singularities of the resolvent and are naturally associated with Jordan blocks and non-exponential dynamics \cite{KatoBook,Weintraub2022,Berry2004,Rotter2009,Heiss2012}. In condensed-matter systems they provide a powerful framework to engineer and classify quantum phase transitions and novel topological phases, enabling robust boundary phenomena 
rooted in the singular topology of the complex energy spectrum \cite{Heiss1990,Cejnar2007,Gonzalez2016,San-Jose2016,Gonzalez2017,Gong2018,Molina2018,Bergholtz2021}. In optics and photonics they underpin striking effects such as enhanced sensitivity and chiral mode transfer, and have motivated extensive experimental and theoretical activity \cite{Klaiman2008,Zhen2015,MiriAlu2019}.

In quantum Markovian dynamics, the generator of evolution is the Liouvillian superoperator of a Lindblad master equation. Liouvillians are intrinsically non-Hermitian, and can host \emph{Liouvillian exceptional points} (LEPs)---degeneracies in the Liouvillian spectrum where eigenmatrices coalesce \cite{Minganti2019}. LEPs have distinct features compared to EPs of effective non-Hermitian Hamiltonians, because the Liouvillian necessarily incorporates quantum noise and jump processes \cite{Minganti2019,Minganti2020Hybrid}. A growing body of work has developed both the mathematical characterization and the physical consequences of LEPs, including explicit constructions of LEPs of higher order and their imprint on coherence and spectral functions \cite{ArkhipovPRA101,ArkhipovPRA102}, as well as recent reviews emphasizing dynamical phenomena such as chiral state transfer upon parametric encircling \cite{SunYi2024Review}. On the experimental side, postselected quantum-trajectory protocols have enabled the observation of EP physics in superconducting circuits \cite{Naghiloo2019NatPhys}, and related approaches have clarified how trajectory conditioning connects Hamiltonian and Liouvillian notions of EPs \cite{Minganti2020Hybrid}.

A central open question is \emph{how} and \emph{when} Liouvillian EP structure becomes visible in experimentally natural observables, in particular in frequency-resolved emission spectra. Unlike traditional Hermitian systems, the spectral behavior near EPs exhibits non-Lorentzian features, including super-Lorentzian line shapes, due to the enhanced non-orthogonality and spectral sensitivity inherent to these degeneracies. Hashemi \emph{et al.} demonstrated that higher-order EPs result in power-law broadened responses surpassing Lorentzian limits, fundamentally shaping the emission profiles of such systems \cite{Hashemi2022}. Extending this, Bid and Schomerus showed how these features persist even beyond the EP itself, emphasizing the non-trivial topology of the system's spectrum \cite{Bid2025}. Simonson {\em et al.} further revealed that quantum noise near EPs leads to a hybrid of Lorentzian and super-Lorentzian spectral features, which can enhance or suppress emission depending on system parameters \cite{Simonson2022}. Kullig {\em et al.} analytically connected these spectral anomalies to a generalized Petermann factor, providing a deeper theoretical basis for emission scaling in EP-governed regimes \cite{Kullig2025}. Together, these studies frame the emission spectrum in the exceptional phase not merely as broadened or shifted, but as a signature of underlying non-Hermitian singularities.

{\color{black} Here, following Ref.~\cite{RubioGarciaPRA106}, we use the term \emph{exceptional spectral phase} to denote a region of the Liouvillian spectrum in which an extensive subset of modes becomes defective in the thermodynamic limit, rather than a situation in which all non-stationary eigenvalues are defective. Related notions of exceptional points and non-diagonalizable Liouvillian dynamics have been discussed in different contexts, particularly for quadratic bosonic systems~\cite{Thompson2023}, where the focus is typically on the structure of the effective dynamical matrix rather than on a spectral phase of the full many-body Liouvillian.}

In this work we address this question in the setting of a dissipative collective spin coupled to a polarized Markovian bath. This model exhibits an \emph{exceptional spectral phase} (ESP) in which, in the thermodynamic limit, an extensive region of the Liouvillian spectrum becomes populated exclusively by second-order EPs, separated from a normal spectral phase by a critical line where the spectral density diverges \cite{RubioGarciaPRA106}. By an exceptional spectral phase (ESP) we mean a regime in which, in the thermodynamic limit, a finite fraction of the Liouvillian spectrum becomes defective, forming an extensive set of second-order exceptional points rather than isolated degeneracies. The same general framework also connects naturally to symmetry-resolved Lindblad dynamics and trajectory physics, where strong symmetries can induce sector selection and dissipative freezing at the level of individual quantum trajectories \cite{SMunozPRA100}.

Here we compute the emission spectrum from the Liouvillian resolvent and demonstrate a state-dependent visibility of exceptional structure. While LEPs are a property of the Liouvillian itself, their spectroscopic signatures depend on the overlap between the emission channel and the defective subspace. We show that steady-state fluorescence can strongly suppress the second-order (super-Lorentzian) contribution even deep in the ESP, whereas spectra sourced from generic initial states (infinite temperature or random full-rank states) reveal a pronounced and statistically significant higher-order pole contribution. This establishes a practical spectroscopic diagnostic of Liouvillian exceptional phases and clarifies why steady-state emission may fail to reveal them even when the underlying Liouvillian is defective.

\section{Model}
We consider a collective spin of length $j$ with Hamiltonian
\begin{equation}
H=-h J_z,
\end{equation}
and Markovian open-system dynamics described by a Lindblad master equation 
$\dot{\rho}=\mathcal{L}\rho$ \cite{Lindblad1976,Gorini1976},
\begin{equation}
\mathcal{L}\rho=-i[H,\rho]+\sum_{\ell\in\{0,+,-\}}
\left(L_\ell\rho L_\ell^\dagger-\tfrac12\{L_\ell^\dagger L_\ell,\rho\}\right).
\label{eq:lindblad}
\end{equation}
The jump operators are
\begin{equation}
L_0=\sqrt{\Gamma_0/j}\,J_z,\qquad
L_\pm=\sqrt{\Gamma(1\mp p)/(2j)}\,J_{\pm},
\label{eq:jumps}
\end{equation}
where $J_\alpha$ are collective $\mathfrak{su}(2)$ generators, $\Gamma$ and $\Gamma_0$
set the dissipative rates, and $p\in[-1,1]$ controls the polarization of the environment. This model was first studied by Riberio and Prosen as an example of exactly solvable many-body Liouvillian \cite{Ribeiro2019}. In Appendix A we discuss the vectorized form of the Liouvillian, the symmtries and the complex spectrum of the model. 

\subsection{Physical realizations}
Equation~\eqref{eq:lindblad} describes a broad class of collective open quantum systems.
A paradigmatic realization is an ensemble of $N=2j$ identical two-level atoms restricted
to the fully symmetric Dicke manifold, where $J_{\pm}=\sum_{k=1}^N\sigma_\pm^{(k)}$ and
$J_z=\tfrac12\sum_{k=1}^N\sigma_z^{(k)}$.
In this context, the operators $L_\pm$ describe collective spontaneous emission and absorption,
while $L_0$ represents collective dephasing due to, e.g., elastic light scattering or
fluctuating fields \cite{Dicke1954,GrossHaroche1982,GardinerZollerQuantumNoise}.
The bath polarization parameter $p$ interpolates between unbiased decay ($p=0$),
corresponding to an infinite-temperature environment, and a fully polarized bath
($p=\pm1$), where either emission or absorption is suppressed and the dynamics becomes
strongly directional. For intermediate values of $p$ and taking into account detailed balance it is possible to relate the jump operators to a bath with inverse temperature $\beta=\frac{1}{h}ln(\frac{1-p}{1+p})$ \cite{RubioGarciaPRA106}. 

Closely related Lindblad structures also arise in solid-state settings.
For example, a localized magnetic moment or quantum dot spin coupled to a spin-polarized
electronic reservoir experiences incoherent spin-flip processes whose rates depend on the
reservoir polarization, together with longitudinal dephasing due to elastic scattering
\cite{Fransson2010,Delgado2010}.
In this interpretation, $J_{\pm}$ generate collective (or effective) spin flips and
$p$ encodes the degree of spin polarization of the leads or substrate.
Such models are directly relevant to scanning tunneling microscopy experiments on single
magnetic atoms and engineered nanostructures, where spin dynamics is probed via
frequency-resolved noise or ESR-type measurements \cite{BaumannScience2015}.

\subsection{Emission spectrum and physical meaning}
We define the emission spectrum associated with collective spin lowering as
\begin{equation}
S(\omega)
=
\lim_{t\to\infty}
\frac{1}{\pi}\,
\operatorname{Re}
\int_0^\infty d\tau\, e^{{\color{black}-}i\omega\tau}
\langle J_+(t)\,J_-(t+\tau)\rangle,
\label{eq:Sdef}
\end{equation}
where the expectation value is taken with respect to the system density matrix at time $t$.
As mentioned earlier, in quantum-optical realizations, $J_-$ plays the role of a collective dipole operator: for an ensemble of $N$ identical two-level emitters in the symmetric manifold, one has $J_-=\sum_{k=1}^N \sigma_-^{(k)}$ and $J_+=J_-^\dagger$.
In the far field, the positive-frequency component of the radiated electric field is proportional to $J_-$ (up to geometry- and polarization-dependent factors), so that $S(\omega)$ corresponds to the frequency-resolved power spectrum of collective fluorescence (resonance fluorescence in the driven case) \cite{GardinerZollerQuantumNoise,CarmichaelSMQO1,Dicke1954,GrossHaroche1982}.
In this sense, $S(\omega)$ is an experimentally accessible observable in platforms ranging from cold-atom ensembles and superradiant emitters to cavity-QED implementations of collective decay \cite{Dicke1954,GrossHaroche1982}.


The same correlator has a natural interpretation in solid-state realizations where a localized spin couples to itinerant electrons or to a detector.
For instance, in ESR-STM measurements of a single magnetic atom on a surface, the detected signal is sensitive to transverse spin dynamics at the Larmor frequency and is commonly expressed in terms of spin correlation functions closely related to $\langle S_+(t) S_-(t+\tau)\rangle$ (or equivalently to the corresponding dynamical susceptibility) \cite{BaumannScience2015,FranssonPRB2010}.
In this context, $S(\omega)$ captures the spectral content of spin-flip fluctuations and can be interpreted as a ``spin-emission'' (or spin-noise) spectrum of the localized moment.%
\footnote{The precise proportionality between the measured current noise / conductance features and spin correlators depends on the microscopic tunneling Hamiltonian and detection protocol, but transverse spin spectra are the central theoretical object in both spin-noise and ESR-STM descriptions \cite{BaumannScience2015,FranssonPRB2010}.}

\subsection{Quantum regression and resolvent form}
Using the quantum regression theorem (QRT) \cite{GardinerZollerQuantumNoise}, Eq.~\eqref{eq:Sdef} becomes
\begin{equation}
S(\omega)
=
\lim_{t\to\infty}
\frac{1}{\pi}\,
\operatorname{Re}
\int_0^\infty d\tau\, e^{{\color{black}-}i\omega\tau}\,
\mathrm{Tr}\left[J_-\,e^{\mathcal{L}\tau}\big(\rho(t)J_+\big)\right].
\label{eq:Sregression}
\end{equation}

{\color{black} \noindent
Equation~\eqref{eq:Sregression} corresponds to the Schrödinger-picture formulation of the quantum regression theorem, in which the operator $J_-$ is kept fixed and the density-matrix-like object $\rho(t)J_+$ is propagated with the Liouvillian.}


For a unique steady state $\rho(t\to\infty)=\rho_{\mathrm{ss}}$, Eq.~\eqref{eq:Sregression} can be written in a resolvent form
\begin{equation}
S(\omega)
=
\frac{1}{\pi}\,
\operatorname{Re}\,
\mathrm{Tr}\left[
J_-\,\big(i\omega-\mathcal{L}\big)^{-1}\big(\rho_{\mathrm{ss}}J_+\big)
\right],
\label{eq:Sresolvent}
\end{equation}
which makes explicit that $S(\omega)$ probes poles of the Liouvillian resolvent.
In particular, defective Liouvillian modes (exceptional points) generate higher-order poles and can therefore produce non-Lorentzian ``super-Lorentzian'' features in $S(\omega)$, provided the source operator $\rho_{\mathrm{ss}}J_+$ has nonzero overlap with the corresponding Jordan subspace.

\subsection{Sector selection rule}
Because $\rho_{\mathrm{ss}}$ is diagonal in the $J_z$ basis, the source operator
$X\equiv\rho_{\mathrm{ss}}J_+$ has matrix elements only of the form
$|m\rangle\langle m-1|$, i.e.\ it lies entirely in the $M=1$ block. Since $\mathcal{L}$
preserves $M$, only Liouvillian modes in sector $M=1$ contribute to $S(\omega)$.

{\color{black}The sector-selection mechanism discussed above is specific to the first-order coherence function underlying the emission spectrum $S(\omega)$, which probes the source operator $X = \rho_{\mathrm{ss}} J^{+}$ and therefore selects the $M=1$ symmetry sector. This raises the question of whether Liouvillian exceptional-point signatures may nevertheless be visible in steady-state observables that probe different sectors. 

In particular, the second-order coherence function $g^{(2)}(\tau)$ involves the operator $Y = J^{-} \rho_{\mathrm{ss}} J^{+}$, which belongs to the $M=0$ sector. In the present model, this sector contains Liouvillian modes that become defective in the thermodynamic limit~\cite{RubioGarciaPRA106}. As a consequence, the time dependence of $g^{(2)}(\tau)$ can in principle exhibit non-exponential decay associated with Jordan-block structure, even when the first-order coherence remains purely exponential. 

This illustrates that the visibility of exceptional-point physics is not solely determined by the Liouvillian spectrum, but also by the overlap between the observable and the corresponding symmetry sector. The present work focuses on the first-order emission spectrum, for which the steady-state contribution is dominated by a non-defective mode, leading to an effectively Lorentzian line shape. Higher-order correlation functions provide a complementary route to access defective sectors and will be explored elsewhere.}

\section{Exceptional-Point Signatures and State-Dependent Visibility}

\subsection{Generalized spectra for arbitrary initial states}
To probe Liouvillian structure beyond steady-state fluorescence, we also consider
the spectrum sourced by an arbitrary state $\rho_0$,
\begin{equation}
S(\omega;\rho_0)
=
\frac{1}{\pi}\,
\operatorname{Re}\,
\textbf{Tr}\left[
J_m\,\big(i\omega-\mathcal{L}\big)^{-1}\big(\rho_0 J_p \big)
\right],
\label{eq:Srho0}
\end{equation}
including $\rho_0=\mathbb{I}/d$ (infinite temperature) and random full-rank states.


\subsection{Lorentzian vs super-Lorentzian discrimination}

Equation~\eqref{eq:Sresolvent} shows that the emission spectrum probes the resolvent
$(i\omega-\mathcal{L})^{-1}$ of the Liouvillian.
If $\mathcal{L}$ is diagonalizable, isolated eigenmodes contribute simple poles of the form
$(\gamma_\mu+i(\omega-\omega_\mu))^{-1}$, yielding Lorentzian line shapes.
By contrast, if $mathcal{L}$ possesses a defective eigenvalue $\lambda_\mu$ associated with
a Jordan block of size two, the resolvent develops a second-order pole
$(i\omega-\lambda_\mu)^{-2}$, producing a qualitatively distinct ``super-Lorentzian''
contribution \cite{KatoBook,Heiss2012,Hashemi2022,Simonson2022}.
Such higher-order poles are the direct spectral signature of Liouvillian exceptional points. More details about the derivation can be found in the Appendix.

Motivated by this structure, we analyze the emission spectrum near its dominant peak using
two nested phenomenological models.
Model~A (Lorentzian) assumes a single simple pole,
\begin{equation}
S_A(\omega)
=
\frac{a}{(\omega-\omega_0)^2+\gamma^2}+c,
\label{eq:modelA}
\end{equation}
while Model~B (Lorentzian + super-Lorentzian) includes an additional second-order pole,
\begin{equation}
S_B(\omega)
=
\frac{a}{(\omega-\omega_0)^2+\gamma^2}
+
\frac{b\,[\gamma^2-(\omega-\omega_0)^2]}
{\big[(\omega-\omega_0)^2+\gamma^2\big]^2}
+c.
\label{eq:modelB}
\end{equation}
Model~B reduces continuously to Model~A for $b=0$ and therefore provides a minimal extension
that captures the effect of a size-two Jordan block in the Liouvillian spectrum.
The parameter $b$ quantifies the relative weight of the second-order pole contribution.

To characterize the importance of the exceptional-point component, we define the EP weight
\begin{equation}
r \equiv \frac{|b|}{|a|+|b|},
\label{eq:rdef}
\end{equation}
which measures the fraction of spectral weight associated with the super-Lorentzian term.
While $r$ is not an order parameter in a strict sense, a nonzero value signals a resolvent
contribution that cannot be generated by diagonalizable Liouvillian modes alone.

Because Model~B contains Model~A as a special case, visual comparison of fits is insufficient
to assess the statistical relevance of the additional parameter.
We therefore employ information-criterion-based model selection.
Specifically, we compute the 
Bayesian information criterion (BIC),
\begin{equation}
\mathrm{BIC}=k\ln N+N\ln(\mathrm{RSS}/N),
\end{equation}
where $k$ is the number of fit parameters, $N$ the number of data points, and
$\mathrm{RSS}$ the residual sum of squares \cite{Schwarz1978,BurnhamAnderson2002}.
Negative values of $\Delta\mathrm{BIC}=\mathrm{BIC}_B-\mathrm{BIC}_A$ indicate that the
super-Lorentzian model is statistically favored despite its larger parameter count.
This criterion provides a robust, quantitative discriminator between ordinary Lorentzian
spectra and genuine exceptional-point-induced line shapes.

%

\section{Numerical Results}

We now present numerical results for the emission spectrum and its exceptional-point
diagnostics in the collective spin model~\eqref{eq:lindblad}–\eqref{eq:jumps}. Numerically, the resolvent is evaluated by explicit construction and inversion of the Liouvillian matrix in the relevant symmetry sector, and fits are performed in a window around the dominant peak.
Unless otherwise stated, we fix $h=1$ and scale the dissipative rates with system size
as $\Gamma\propto j^0$, which yields a nontrivial thermodynamic limit \cite{RubioGarciaPRA106}.
All spectra are computed from the Liouvillian resolvent~\eqref{eq:Sresolvent} and analyzed
using the Lorentzian and super-Lorentzian models introduced in
Sec.~III.
{\color{black} At finite system size, the Liouvillian remains strictly diagonalizable, and the emission spectrum can formally be expressed as a sum of Lorentzian contributions associated with distinct eigenmodes. Therefore, strictly speaking, no exact super-Lorentzian lineshape arises at finite size. 

However, as shown in Fig.~2(b) of Ref.~\cite{RubioGarciaPRA106}, the eigenvector distance between pairs of Liouvillian modes decreases rapidly with system size, exhibiting an exponential scaling toward coalescence. As a consequence, already for moderate values of $j$, pairs of modes become nearly degenerate both in their eigenvalues and eigenvectors. In this regime, the corresponding Lorentzian contributions strongly overlap and cannot be resolved individually.

The resulting lineshape is therefore effectively described by a Lorentzian plus super-Lorentzian form, which captures the leading correction associated with the near-coalescence of modes. In the time domain, this corresponds to dynamics approaching the $(a + b\tau)e^{-\gamma \tau}$ behavior characteristic of second-order exceptional points in the thermodynamic limit. We thus interpret the observed non-Lorentzian features as finite-size precursors of the exceptional spectral phase. A quantitative discussion can be found in Appendix A.8.
}
\subsection{Steady-state versus generic-state emission}
\begin{figure}[t]
  \centering
  \begin{subfigure}{\linewidth}
    \centering
    \includegraphics[width=\linewidth]{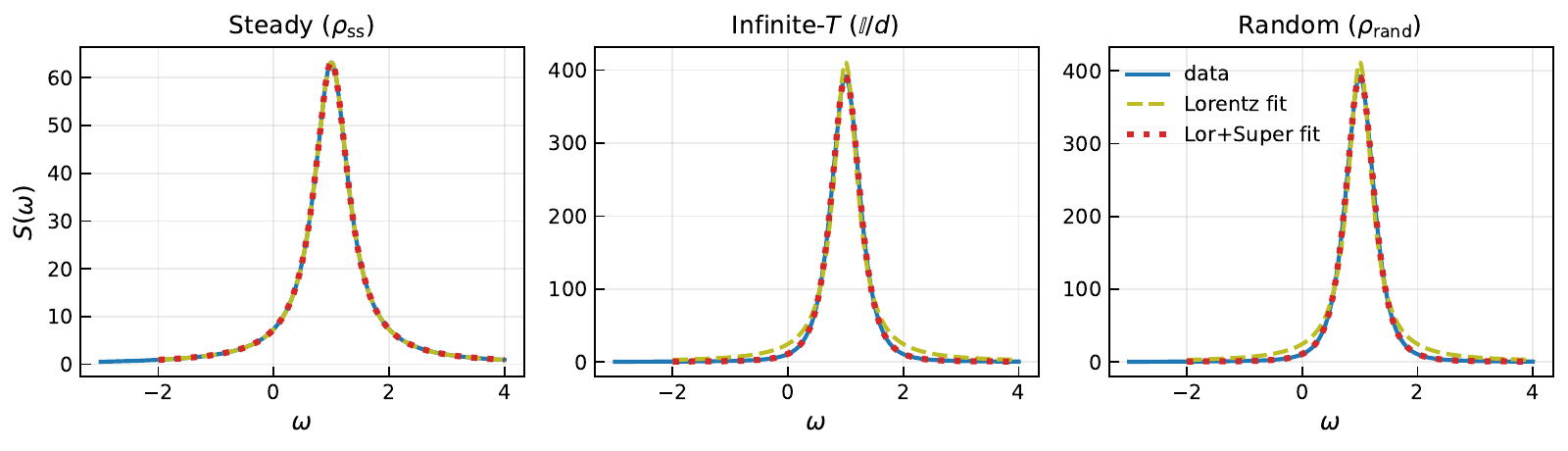}
    \caption{$p=0.2$}
    \label{fig:Somega_p02}
  \end{subfigure}

  \vspace{0.4cm}

  \begin{subfigure}{\linewidth}
    \centering
    \includegraphics[width=\linewidth]{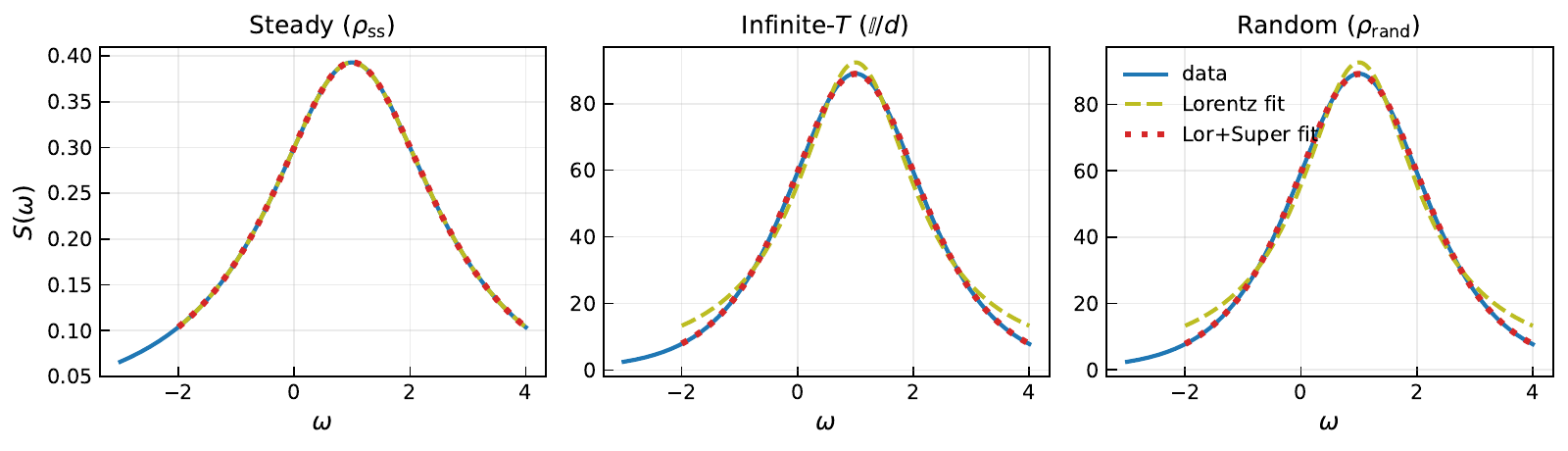}
    \caption{$p=0.9$}
    \label{fig:Somega_p09}
  \end{subfigure}

  \caption{
    Emission spectra $S(\omega)$ computed from the Liouvillian resolvent for two representative values of the {\color{black} polarization parameter} $p$.
    In each panel, spectra are shown for the steady state $\rho_{\mathrm{ss}}$, the infinite-temperature state $\rho_{\infty}=\mathbb{1}/d$, and a random reference state, together with fits to a single Lorentzian and to a Lorentzian-plus-superlorentzian model.
    For weak pumping ($p=0.2$, upper panel), 
    {\color{black} Steady state spectra are well described by a single lorentzian for both values of $p$, indicating isolated Liouvillian resonances. However at stronger pumping ($p=0.9$, lower panel), the infinite-temperature and random-state spectra develop clearer deviations from a single-lorentzian lineshape and are well fitted by a Lorentzian plus super-lorentzian lineshape.}    
    This behavior reflects the increasing role of collective and near-degenerate Liouvillian modes at large $p$.
  }
  \label{fig:Somega_two_p}
\end{figure}

Figures~\ref{fig:Somega_p02} and~\ref{fig:Somega_p09} compare the emission spectrum obtained from the steady state
$\rho_{\mathrm{ss}}$ with that obtained from a generic initial random state and an
infinite-temperature state $\rho_0=\mathbb{I}/d$, for identical Liouvillian parameters with two values of the polarization parameter $p=0.2$ and $p=0.9$.
In both cases the spectrum exhibits a dominant peak at $\omega = h$, whose width
and position are controlled by the decay rates and coherent precession frequency. {\color{black} Before turning to the numerical analysis, it is instructive to consider the exactly solvable limits $p=0$ and $p=1$, where explicit analytical expressions for the first-order coherence function and the emission spectrum can be obtained. As shown in Appendix~\ref{app:analytic_limits}, in both cases the steady-state spectrum is exactly Lorentzian, despite the qualitatively different structure of the Liouvillian spectrum in each limit. We note, however, that this absence of super-Lorentzian features is specific to the steady-state observable considered here. For more general initial states, the source operator can overlap with defective Liouvillian modes, and the corresponding resolvent may contain genuine second-order pole contributions. This is illustrated explicitly in Appendix~\ref{app:p1_generic} for the exactly solvable case $p=1$, where the Liouvillian is maximally exceptional but the steady-state spectrum remains purely Lorentzian. These results provide useful benchmarks and help clarify the role of symmetry-sector selection discussed below.}

Figure~\ref{fig:Somega_two_p} illustrates how the structure of the emission spectrum evolves as the incoherent pumping strength {\color{black} bath polarization} is increased.
For moderate pumping ($p=0.2$), shown in the upper panel, the spectra obtained from the steady state, the infinite-temperature state, and a random reference state are all well captured by a single Lorentzian profile, although with small deviations in the infinite temperature and random initial states.
This indicates that the Liouvillian spectrum is dominated by well-isolated decay modes, and that the spectral response is largely insensitive to the choice of initial state.

In contrast, for strong pumping ($p=0.9$), shown in the lower panel, 
{\color{black} the infinite-temperature and random-state spectra exhibit} pronounced deviations from a single-Lorentzian form and {\color{black}are } better described by a model that includes an additional superlorentzian contribution.
This qualitative change is absent in the 
{\color{black} steady state}
which remain comparatively smooth.
The enhanced sensitivity of the steady-state response at large $p$ is consistent with the appearance of near-degenerate Liouvillian modes and the spectral crowding discussed in the context of Fig.~\ref{fig:appendix_liouvillian_spectrum_vectorization} {\color{black} in the Appendix}.
Together, these results highlight how collective dissipation and Liouvillian spectral structure manifest directly in experimentally accessible frequency-domain observables. 
{\color{black} For both values of the parameter $p$}
the infinite temperature and the random initial states have almost the same emission spectrum.

Despite the presence of an exceptional spectral phase (ESP) in the Liouvillian spectrum,
the steady-state emission is accurately described by a purely Lorentzian line shape.
Information-criterion analysis strongly favors the Lorentzian model, with large positive
values of $\Delta\mathrm{BIC}$, and the extracted EP weight $r$ remains close to zero.
By contrast, for the infinite-temperature state and the random states the Lorentzian model fails to capture the
line shape: the inclusion of a super-Lorentzian term yields a dramatic reduction of the
residual error and is overwhelmingly favored by both AIC and BIC.
The corresponding EP weight reaches values of order $r\sim 10^{-1}$, providing clear
evidence for a second-order pole contribution associated with a defective Liouvillian mode.

This striking contrast demonstrates that exceptional points are a property of the
Liouvillian generator, but their spectroscopic visibility depends crucially on the
overlap between the emission channel and the defective subspace.
In the present model, the steady state becomes increasingly polarized as $p\to 1$,
which suppresses the source operator $\rho_{\mathrm{ss}}J_+$ in the $M=1$ sector and effectively filters
out the exceptional contribution, even though the Liouvillian itself remains defective. {\color{black} Although both the steady-state density matrix and the infinite-temperature state are diagonal in the $J_z$ basis, they induce very different source operators $X=\rho J^+$ in the $M=1$ sector. The steady state is strongly polarized for $p>0$, so $X=\rho_{\mathrm{ss}}J^+$ is concentrated near the edge of the ladder and predominantly overlaps with the non-defective mode of largest real part. As a result, the second-order-pole contribution is strongly suppressed and the spectrum remains effectively Lorentzian. By contrast, the infinite-temperature state weights all $J_z$ sectors uniformly, so that $X=(\mathbb I/d)J^+$ has broad support across the full $M=1$ block, including the near-defective modes associated with the exceptional spectral phase. This broader overlap makes the super-Lorentzian contribution visible in the corresponding spectrum.}

\subsection{Exceptional-point diagnostics versus bath polarization}
\begin{figure}[t]
\centering
\includegraphics[width=\linewidth]{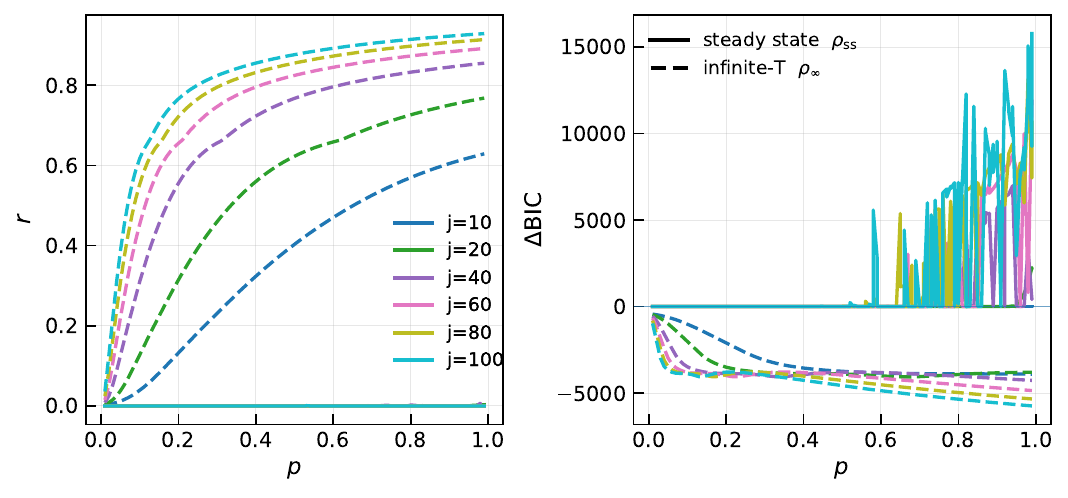}
\caption{
{\color{black} \textbf{Liouvillian exceptional-point diagnostics.}
Extracted EP weight $r$ as a function of the polarization parameter $p$ for different values of the system size $j$ (left panel) and information-criterion difference $\Delta\mathrm{BIC}$ (right panel).}
}
\label{fig:main}
\end{figure}

To characterize the emergence of exceptional behavior systematically, we analyze the
EP diagnostics as a function of bath polarization $p$ and system size $j$. 
Figure~\ref{fig:main} shows the EP weight $r(p)$ ({\color{black} left}) and the model-selection indicator
$\Delta\mathrm{BIC}(p)$ ({\color{black} right}) extracted from spectra computed with infinite-temperature
and steady-state sources.

For generic initial states, both diagnostics reveal a sharp crossover as $p$ is increased:
$r(p)$ grows rapidly from zero and $\Delta\mathrm{BIC}$ becomes strongly negative,
signaling the onset of a statistically significant super-Lorentzian contribution.
As the system size increases, this crossover becomes increasingly sharp and shifts toward
smaller values of $p$, consistent with the approach to the thermodynamic exceptional
spectral phase identified in Ref.~\cite{RubioGarciaPRA106}.
By contrast, the same diagnostics applied to steady-state spectra remain near their
Lorentzian values for all $p$, reflecting the steady-state filtering discussed above.



This analysis clarifies that the EP weight $r$ is not an order parameter, but rather a
measure of spectral dominance: it quantifies when defective modes control the observable
response, not merely when they exist.
Consequently, the boundary of the exceptional spectral phase is more robustly identified
by information-criterion diagnostics than by the maximum of $r(p)$.

\section{Discussion and Outlook}

We have shown that exceptional points of a Liouvillian generator can leave clear and
quantifiable fingerprints in emission spectra, provided that the emission channel has
sufficient overlap with the defective subspace.
In a dissipative collective spin model exhibiting an exceptional spectral phase, we
identified super-Lorentzian line shapes as the direct manifestation of second-order poles
in the Liouvillian resolvent.
Crucially, we demonstrated that the visibility of these signatures is strongly
state-dependent: while generic initial states reveal the exceptional structure through
statistically significant deviations from Lorentzian behavior, steady-state emission can
remain essentially blind to it due to polarization-induced filtering.

This state-dependent visibility clarifies an important conceptual point.
Exceptional points are properties of the dynamical generator, not of a particular
observable or preparation, and their experimental detectability depends on how the system
is probed.
In the present model, the steady state suppresses the source operator that couples to the
defective modes, even deep inside the exceptional spectral phase.
Our results therefore explain why steady-state fluorescence measurements may fail to
detect Liouvillian exceptional points, and why transient, quenched, or otherwise
nonequilibrium protocols are often essential.

The spectroscopic diagnostics introduced here—super-Lorentzian fitting combined with
information-criterion-based model selection—provide a practical and broadly applicable
toolkit for identifying Liouvillian exceptional behavior.
Because these diagnostics rely only on frequency-resolved correlation functions, they are
directly relevant to a wide range of experimental platforms, including collective atomic
ensembles, cavity and circuit QED systems, and nanoscale magnetic structures probed by
spin-resolved spectroscopy.
In particular, our analysis suggests that measurements performed after a quench or from
engineered mixed states may be more sensitive to exceptional physics than steady-state
fluorescence alone.

Several natural extensions follow from this work.
First, it would be interesting to generalize the present analysis to higher-order
Liouvillian exceptional points, where the resolvent contains poles of order three or
higher and the resulting spectral line shapes are expected to be even more distinctive.
Second, the role of individual quantum trajectories merits further investigation:
trajectory-conditioned dynamics are known to exhibit exceptional points even when the
unconditional Liouvillian does not, and conversely may provide enhanced sensitivity to
defective modes.
Finally, extending these ideas to spatially extended or multimode systems could establish
a direct link between exceptional spectral phases and transport or many-body localization
phenomena in open quantum matter.

More broadly, our results highlight emission spectroscopy as a powerful probe of the
non-Hermitian structure of open quantum systems.
By explicitly connecting Liouvillian exceptional points to observable spectral features,
we provide a concrete route for exploring non-Hermitian criticality in realistic quantum
platforms. {\color{black} In particular, observables probing different symmetry sectors, such as higher-order correlation functions, may provide access to defective Liouvillian modes even in the steady state.}




\paragraph{Funding information}
This work has been supported by the Agencia Estatal de Investigación from
Spain (Grant PID2022-136285NB-C31).


\begin{appendix}
\numberwithin{equation}{section}


\section{Liouville-space formulation and weak symmetry}

\subsection{Vectorization of the Liouvillian}

To analyze the spectral properties of the Liouvillian appearing in
Eqs.~(1)–(3) of the main text, it is convenient to work in Liouville
space by vectorizing the density matrix \cite{AmShallem2015}.
We adopt the standard column-stacking convention,
\begin{equation}
\mathrm{vec}(\rho)
=
\sum_{m,n}\rho_{mn}\,|m\rangle\otimes|n\rangle,
\end{equation}
under which operator multiplication maps as
\begin{equation}
\mathrm{vec}(A\rho B)
=
(A\otimes B^{\mathsf T})\,\mathrm{vec}(\rho).
\end{equation}

Using this representation, the Lindblad generator
\begin{equation}
\mathcal{L}\rho
=
-i[H,\rho]
+
\sum_\ell
\left(
L_\ell\rho L_\ell^\dagger
-\tfrac12\{L_\ell^\dagger L_\ell,\rho\}
\right)
\end{equation}
is mapped onto a non-Hermitian superoperator matrix
\begin{equation}
\mathcal{L}
=
-i\bigl(H\otimes\mathbb{I}-\mathbb{I}\otimes H^{\mathsf T}\bigr)
+
\sum_\ell
\left(
L_\ell\otimes L_\ell^*
-\tfrac12 L_\ell^\dagger L_\ell\otimes\mathbb{I}
-\tfrac12\mathbb{I}\otimes(L_\ell^\dagger L_\ell)^{\mathsf T}
\right),
\label{eq:Liouville_matrix}
\end{equation}
acting on the Liouville-space vector
$|\rho\rangle\rangle=\mathrm{vec}(\rho)$.
The Liouvillian spectrum is obtained from
\begin{equation}
\mathcal{L}|\rho_\mu\rangle\rangle
=
\lambda_\mu|\rho_\mu\rangle\rangle,
\end{equation}
where the eigenvalues $\lambda_\mu$ are generally complex and may form
nontrivial Jordan blocks.

\subsection{Explicit vectorized Liouvillian for the collective-spin model}

For completeness, we write the Liouvillian matrix $\mathcal{L}$ explicitly in the
vectorized representation for the model of Eqs.~\eqref{eq:lindblad}–\eqref{eq:jumps}.
Following Refs.~\cite{RubioGarciaPRA106}, we map an $N\times N$ density matrix
$\rho_{\alpha,\beta}$ (with $N=2j+1$) to a vector $|\alpha,\beta\rangle\rangle$ in a
Hilbert space of dimension $N^2$.
Left and right multiplications by collective-spin operators are represented as
\begin{equation}
J\,\rho \;\mapsto\; (J\otimes \mathbb{I})|\rho\rangle\rangle \equiv K_1\,|\rho\rangle\rangle,
\qquad
\rho\,J \;\mapsto\; (\mathbb{I}\otimes J^{\mathsf T})|\rho\rangle\rangle \equiv K_2\,|\rho\rangle\rangle,
\end{equation}
so that $K_{1\alpha}$ and $K_{2\alpha}$ ($\alpha=x,y,z$) form two commuting spin-$j$
$\mathfrak{su}(2)$ algebras acting on the left and right indices, respectively.

Using $H=-hJ_z$ and the jumps
$L_0=\sqrt{\Gamma_0/j}\,J_z$ and
$L_\pm=\sqrt{\Gamma(1\mp p)/(2j)}\,J_\pm$,
the Liouvillian superoperator $\mathcal{L}$ becomes an explicit non-Hermitian operator
on Liouville space:
\begin{align}
\mathcal{L}
={}&
-\Gamma\,(j+1)
+i h\,(K_{1z}-K_{2z})
+\frac{\Gamma}{j}\,K_{1z}K_{2z}
+\frac{\Gamma-\Gamma_0}{2j}\,(K_{1z}-K_{2z})^2
\nonumber\\
&\quad
-\frac{\Gamma}{j}\,\frac{p}{2}\,(K_{1z}+K_{2z})
+\frac{\Gamma}{j}\,\frac{1-p}{2}\,K_{1+}K_{2+}
+\frac{\Gamma}{j}\,\frac{1+p}{2}\,K_{1-}K_{2-}.
\label{eq:L_vectorized_explicit}
\end{align}
Equation~\eqref{eq:L_vectorized_explicit} is the form used in
Ref.~\cite{RubioGarciaPRA106} (their Eq.~(4)), expressed in our notation.
The $z$ component
\begin{equation}
K_z \equiv K_{1z}-K_{2z}
\end{equation}
commutes with $\mathcal{L}$, $[\mathcal{L},K_z]=0$, and thus generates a weak symmetry
that decomposes Liouville space into independent sectors labeled by
$M\in\{-2j,-2j+1,\ldots,2j\}$, corresponding to fixed eigenvalues of $K_z$
\cite{RubioGarciaPRA106,BuccaPRL2018,SMunozPRA100}.
The term proportional to $\Gamma_0$ produces a sector-dependent real shift,
since $(K_{1z}-K_{2z})$ has fixed eigenvalue $M$ within a given weak-symmetry block.

\subsection{Liouvillian spectrum and near-degeneracies}
\label{app:liouvillian-spectrum}

\begin{figure}[t]
  \centering
  \includegraphics[width=\linewidth]{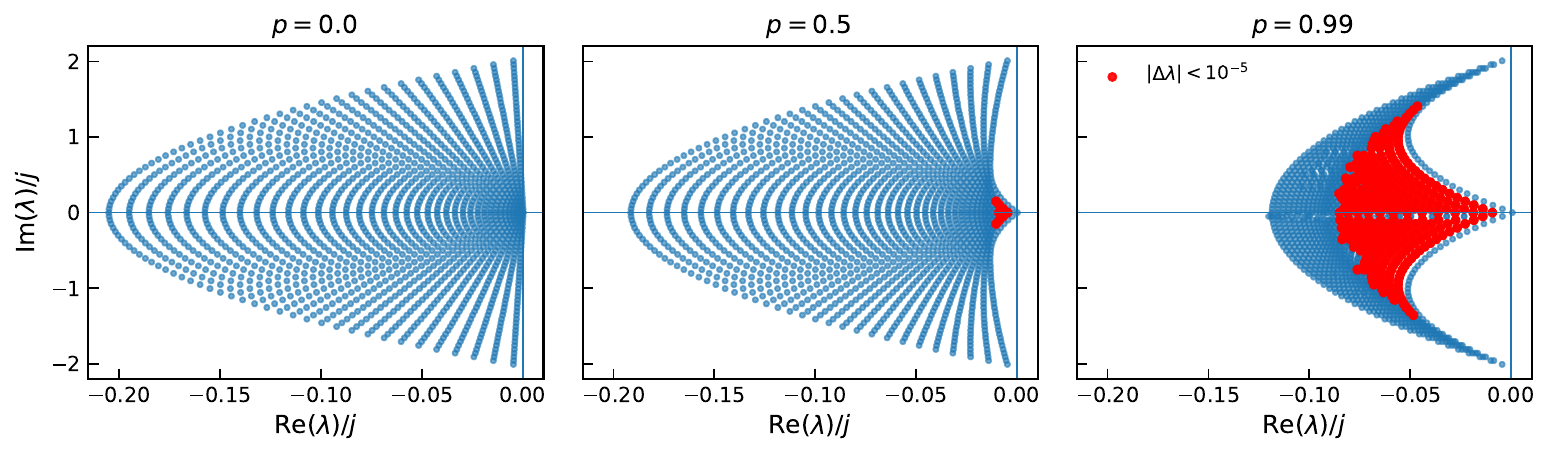}
  \caption{
    Liouvillian spectrum for a collective spin system with $j=20$, obtained by explicit vectorization of the master equation and direct diagonalization of the Liouvillian superoperator.
    The real and imaginary parts of the eigenvalues $\lambda$ are shown (rescaled by $j$) for three representative values of the incoherent pumping parameter: $p=0$, $p=0.5$, and $p=0.99$.
    All eigenvalues are plotted as black points, while eigenvalues that have at least one neighbor in the complex plane closer than $|\Delta\lambda|<10^{-6}$ are highlighted in red.
    The latter provide a numerical diagnostic for near-degeneracies of the Liouvillian spectrum, consistent with the emergence of exceptional-point--like behavior as $p$ is increased.
    Parameters are $h=1$, $\Gamma=0.1$, and $\Gamma_0=0$.
  }
  \label{fig:appendix_liouvillian_spectrum_vectorization}
\end{figure}

Using the vectorized representation of the master equation introduced above, the full Liouvillian superoperator can be constructed explicitly and diagonalized numerically for moderate system sizes.
Figure~\ref{fig:appendix_liouvillian_spectrum_vectorization} shows the resulting Liouvillian spectrum for a collective spin with $j=20$, plotted in the complex plane for several representative values of the bath polarization parameter $p$.

For no bath polarization ($p=0$), the spectrum is well separated, with no visible clustering of eigenvalues.
As $p$ is increased, groups of eigenvalues approach each other in the complex plane in the region close to the steady state, these nearly degenerate region becomes larger for higher values of $|p|$ and is separated from the rest of the spectrum by a region of high density of eigenvalues \cite{Ferreira2019,RubioGarciaPRA106}.
To highlight this behavior, eigenvalues that have at least one neighbor closer than $|\Delta\lambda|<10^{-6}$ are marked in red.
These near-degeneracies provide a clear numerical signature of spectral coalescence in the Liouvillian, consistent with the behavior expected in the vicinity of exceptional points in non-Hermitian generators. Rubio-Garcia {\em et al.} \cite{RubioGarciaPRA106} demonstrated that in the Thermodynamic Limit $j \rightarrow \infty$ the eigenvalues really coalescence in pairs forming what they called an exceptional spectral phase separated with the normal phase of non-degenerate eigenvalues by a  Liouvillian spectral phase transition with a divergent density of states, which is analogous to the excited state phase transitions of closed quantum systems \cite{Cejnar2021}.

\subsection{Sector selection by emission observables}

The emission spectrum studied in the main text involves the source
operator
\begin{equation}
X=\rho\,J_+,
\end{equation}
which in Liouville space contains only basis elements of the form
$|m\rangle\langle m-1|$.
These operators belong exclusively to the sector $M=1$.
Since the Liouvillian does not couple different $M$ sectors, only the
block $\mathcal{L}^{(M=1)}$ contributes to the resolvent expression for
the emission spectrum.

This sector selection explains why exceptional points present in other
symmetry sectors do not influence the observed spectrum, and why the
visibility of exceptional behavior depends sensitively on the structure
of the source state $\rho$.
In particular, steady states that suppress the relevant sector
components can effectively filter out exceptional contributions even
when the Liouvillian itself remains defective.

\subsection{Super-Lorentzian line shape from a Jordan-block resolvent}
\label{SM:JordanSuperLorentz}

In this section we show explicitly why a \emph{defective} Liouvillian (Jordan block)
produces a ``super-Lorentzian'' contribution to the emission spectrum.
We start from the resolvent form of the spectrum in the steady state,
\begin{equation}
S(\omega)
=
\frac{1}{\pi}\,\Re\,
\mathbf{Tr}\left[
J_-\,(i\omega-\mathcal{L})^{-1}\big(\rho_{\rm ss}J_+\big)
\right],
\label{SM:eq:S_resolvent}
\end{equation}
and use the spectral structure of $(i\omega-\mathcal{L})^{-1}$ near an exceptional point.

\paragraph{Jordan block and resolvent.}
Consider an invariant subspace where $\mathcal{L}$ has a size-two Jordan block at
eigenvalue $\lambda_0$,
\begin{equation}
\mathcal{L}\big|_{\rm sub}
=
\lambda_0\,\mathbb{I}
+
N,
\qquad
N=
\begin{pmatrix}
0 & 1\\
0 & 0
\end{pmatrix},
\qquad
N^2=0.
\label{SM:eq:JordanBlock}
\end{equation}
Then, for $z\equiv i\omega$, the resolvent restricted to this subspace can be inverted
exactly:
\begin{align}
(z-\mathcal{L})^{-1}
&=
\big[(z-\lambda_0)\mathbb{I}-N\big]^{-1}
\nonumber\\
&=
\frac{1}{z-\lambda_0}\,
\left[\mathbb{I}-\frac{N}{z-\lambda_0}\right]^{-1}
=
\frac{1}{z-\lambda_0}\,
\left[\mathbb{I}+\frac{N}{z-\lambda_0}\right]
\nonumber\\
&=
\frac{\mathbb{I}}{z-\lambda_0}
+
\frac{N}{(z-\lambda_0)^2}.
\label{SM:eq:ResolventJordan}
\end{align}
Equation~\eqref{SM:eq:ResolventJordan} is the defining analytic signature of a Jordan block:
a \emph{second-order pole} at $z=\lambda_0$ in addition to the usual simple pole.

\paragraph{Projection of the correlation function onto the Jordan chain.}
Let the ``source'' and ``detector'' vectors in Liouville space be
\begin{equation}
|B\rangle\rangle \equiv |\rho_{\rm ss}J_+\rangle\rangle,
\qquad
\langle\langle A| \equiv \langle\langle J_-|,
\label{SM:eq:ABdefs}
\end{equation}
so that Eq.~\eqref{SM:eq:S_resolvent} becomes
\begin{equation}
S(\omega)=\frac{1}{\pi}\Re\;\langle\langle A|(i\omega-\mathcal{L})^{-1}|B\rangle\rangle.
\label{SM:eq:S_AB}
\end{equation}
Assume the Jordan block is spanned by a generalized right eigenbasis
$\{|r_0\rangle\rangle,|r_1\rangle\rangle\}$ and left eigenbasis
$\{\langle\langle \ell_0|,\langle\langle \ell_1|\}$ satisfying
\begin{align}
\mathcal{L}|r_0\rangle\rangle &= \lambda_0 |r_0\rangle\rangle,
&
\mathcal{L}|r_1\rangle\rangle &= \lambda_0 |r_1\rangle\rangle + |r_0\rangle\rangle,
\nonumber\\
\langle\langle \ell_0|\mathcal{L} &= \lambda_0 \langle\langle \ell_0|,
&
\langle\langle \ell_1|\mathcal{L} &= \lambda_0 \langle\langle \ell_1| + \langle\langle \ell_0|.
\label{SM:eq:JordanChains}
\end{align}
A convenient normalization is
$\langle\langle \ell_i|r_j\rangle\rangle=\delta_{ij}$ for $i,j\in\{0,1\}$.
In this basis one may represent
\begin{equation}
\mathbb{I}_{\rm sub} = |r_0\rangle\rangle\langle\langle \ell_0|
+|r_1\rangle\rangle\langle\langle \ell_1|,
\qquad
N = |r_0\rangle\rangle\langle\langle \ell_1|.
\label{SM:eq:IandN_projectors}
\end{equation}
Inserting Eq.~\eqref{SM:eq:ResolventJordan} into Eq.~\eqref{SM:eq:S_AB} and retaining only
the Jordan-block contribution yields
\begin{equation}
\langle\langle A|(i\omega-\mathcal{L})^{-1}|B\rangle\rangle
\simeq
\frac{\alpha}{i\omega-\lambda_0}
+
\frac{\beta}{(i\omega-\lambda_0)^2},
\label{SM:eq:alpha_beta}
\end{equation}
with coefficients determined by overlaps with the Jordan chain,
\begin{equation}
\alpha
=
\langle\langle A|r_0\rangle\rangle\langle\langle \ell_0|B\rangle\rangle
+
\langle\langle A|r_1\rangle\rangle\langle\langle \ell_1|B\rangle\rangle,
\qquad
\beta
=
\langle\langle A|r_0\rangle\rangle\langle\langle \ell_1|B\rangle\rangle.
\label{SM:eq:alpha_beta_overlaps}
\end{equation}
Thus, the emission spectrum near the defective mode is controlled by a simple-pole
term ($\alpha$) and a second-order-pole term ($\beta$).
Importantly, \emph{both poles occur at the same complex frequency} $\lambda_0$, hence
they share a common linewidth.

\paragraph{Real-frequency line shape: Lorentzian + super-Lorentzian.}
Write the defective eigenvalue as
\begin{equation}
\lambda_0 = -\gamma + i\omega_0,
\qquad
\gamma>0,
\label{SM:eq:lambda_param}
\end{equation}
and define $\Delta\equiv \omega-\omega_0$.
Then
\begin{equation}
i\omega-\lambda_0 = \gamma + i\Delta.
\label{SM:eq:denom}
\end{equation}
Using
\begin{equation}
\Re\frac{1}{\gamma+i\Delta}=\frac{\gamma}{\Delta^2+\gamma^2},
\qquad
\Re\frac{1}{(\gamma+i\Delta)^2}=\frac{\gamma^2-\Delta^2}{(\Delta^2+\gamma^2)^2},
\label{SM:eq:Re_identities}
\end{equation}
Eq.~\eqref{SM:eq:alpha_beta} gives the standard decomposition
\begin{equation}
S(\omega)
\simeq
\frac{1}{\pi}\left[
a\,\frac{1}{\Delta^2+\gamma^2}
+
b\,\frac{\gamma^2-\Delta^2}{(\Delta^2+\gamma^2)^2}
\right],
\label{SM:eq:SB_shape}
\end{equation}
where the real amplitudes $a$ and $b$ depend on the overlaps
in Eq.~\eqref{SM:eq:alpha_beta_overlaps}. (Any overall factor of $\gamma$ can be absorbed
into $a$; we adopt the convention used in the main text.)
The first term is a Lorentzian peak (simple pole), while the second is the
\emph{super-Lorentzian} contribution (second-order pole), which is the analytic fingerprint
of a Jordan block.



{\color{black}
\subsection{Analytical expressions in the exactly solvable limits $p=0$ and $p=1$}
\label{app:analytic_limits}

In this appendix we derive explicit analytical expressions for the first-order coherence function and the corresponding emission spectrum in the two limits of zero and full bath polarization, $p=0$ and $p=1$. These limits are exactly solvable and provide useful benchmarks for the general discussion in the main text.

The steady-state emission spectrum is defined as
\begin{equation}
S(\omega)=\frac{1}{\pi}\,\mathrm{Re}\,\mathrm{Tr}\!\left[
J^- (i\omega-\mathcal L)^{-1}(\rho_{\mathrm{ss}}J^+)
\right],
\end{equation}
which is the Fourier transform of the first-order coherence function
\begin{equation}
C^{(1)}(\tau)=\langle J^+(0)J^-(\tau)\rangle_{\mathrm{ss}}
=\mathrm{Tr}\!\left[J^- e^{\mathcal L \tau}(\rho_{\mathrm{ss}}J^+)\right].
\end{equation}

\paragraph*{Case $p=1$.}

For a fully polarized bath ($p=1$), the dissipative dynamics is governed solely by spin-lowering processes, and the steady state is the fully polarized ground state,
\begin{equation}
\rho_{\mathrm{ss}} = |-j\rangle\langle -j|.
\end{equation}
The source operator entering the correlation function is therefore
\begin{equation}
X = \rho_{\mathrm{ss}} J^+ = \sqrt{2j}\,|-j\rangle\langle -j+1|.
\end{equation}

Using the exact Liouvillian spectrum in this limit, which is given by the diagonal elements of $\mathcal L$ in the basis $|m,m-M\rangle$~\cite{RubioGarciaPRA106}, one finds that $X$ is an eigenoperator of the Liouvillian with eigenvalue
\begin{equation}
\lambda_X = ih - \Gamma.
\end{equation}
This eigenvalue corresponds to the mode with largest real part in the symmetry sector $M=1$, which remains non-defective even in the presence of extensive pairwise degeneracies in the rest of the spectrum~\cite{RubioGarciaPRA106}.

As a consequence, the time evolution of the source operator is purely exponential,
\begin{equation}
e^{\mathcal L \tau} X = e^{(ih-\Gamma)\tau} X,
\end{equation}
and the first-order coherence function reads
\begin{equation}
C^{(1)}(\tau) = 2j\, e^{(ih-\Gamma)\tau}.
\end{equation}

The corresponding emission spectrum is therefore exactly Lorentzian,
\begin{equation}
S_{p=1}(\omega) =
\frac{2j}{\pi}\,
\frac{\Gamma}{(\omega-h)^2+\Gamma^2}.
\end{equation}

\paragraph*{Case $p=0$.}

For an unpolarized bath ($p=0$), the model is exactly solvable due to an enhanced $O(3)$ symmetry, and the Liouvillian eigenvalues are given by~\cite{RubioGarciaPRA106}
\begin{equation}
\lambda_{K,M} = ihM - \frac{\Gamma}{2j}K(K+1) + \frac{\Gamma}{2j}M^2,
\end{equation}
where $K$ and $M$ are angular-momentum quantum numbers. In this case, the steady state is the infinite-temperature state,
\begin{equation}
\rho_{\mathrm{ss}} = \frac{\mathbb I}{2j+1}.
\end{equation}

The source operator becomes
\begin{equation}
X = \rho_{\mathrm{ss}} J^+ = \frac{1}{2j+1} J^+.
\end{equation}
Since $J^+$ transforms as a rank-one spherical tensor, it belongs entirely to the $(K,M)=(1,1)$ sector of the Liouvillian. Therefore, it is an eigenoperator with eigenvalue
\begin{equation}
\lambda_{1,1} = ih - \frac{\Gamma}{2j}.
\end{equation}

The time evolution is again purely exponential,
\begin{equation}
e^{\mathcal L \tau} J^+ = e^{(ih-\Gamma/(2j))\tau} J^+,
\end{equation}
leading to
\begin{equation}
C^{(1)}(\tau) =
\frac{1}{2j+1}\,\mathrm{Tr}(J^- J^+)\, e^{(ih-\Gamma/(2j))\tau}.
\end{equation}
Using $\mathrm{Tr}(J^-J^+) = \frac{2}{3}(2j+1)j(j+1)$, one obtains
\begin{equation}
C^{(1)}(\tau) = \frac{2}{3}j(j+1)\, e^{(ih-\Gamma/(2j))\tau}.
\end{equation}

The emission spectrum is therefore
\begin{equation}
S_{p=0}(\omega) =
\frac{1}{\pi}\,
\frac{\frac{2}{3}j(j+1)\,\Gamma/(2j)}
{(\omega-h)^2 + \bigl(\Gamma/(2j)\bigr)^2}.
\end{equation}


In both solvable limits $p=0$ and $p=1$, the steady-state first-order coherence function is a single exponential, leading to an exactly Lorentzian emission spectrum. However, the underlying physical mechanisms are fundamentally different. For $p=1$, the Liouvillian is highly non-diagonalizable and exhibits an extensive set of second-order exceptional points, yet the source operator $\rho_{\mathrm{ss}}J^+$ couples only to the extremal non-defective mode of the $M=1$ sector. For $p=0$, by contrast, the Liouvillian is fully diagonalizable and the operator $J^+$ lies entirely within a single irreducible sector of the exact $O(3)$ solution.

These results provide analytical support for the central conclusion of the main text: the visibility of exceptional-point physics in emission spectra depends not only on the Liouvillian spectrum, but also on the overlap between the observable and the relevant symmetry sector.
}

{\color{black}
\subsection{Second-order pole contributions at $p=1$ for generic initial states}
\label{app:p1_generic}

In this appendix we analyze the structure of the emission spectrum for general initial states in the exactly solvable limit of full bath polarization, $p=1$. In contrast to the steady-state case discussed in the main text, we show that generic initial conditions can give rise to genuine second-order pole contributions in the resolvent, reflecting the exceptional structure of the Liouvillian in this limit.

We consider the generalized emission spectrum
\begin{equation}
S(\omega;\rho_0)
=
\frac{1}{\pi}\,\mathrm{Re}\,\mathrm{Tr}\!\left[
J^- (i\omega-\mathcal L)^{-1}(\rho_0 J^+)
\right],
\end{equation}
where $\rho_0$ is an arbitrary initial state. As discussed in the main text, only the diagonal part of $\rho_0$ in the $J_z$ basis contributes, so we write
\begin{equation}
\rho_0 = \sum_{n=-j}^{j} p_n |n\rangle\langle n|.
\end{equation}

\paragraph*{Triangular structure of the $M=1$ sector.}

At $p=1$, the Liouvillian is exactly solvable and assumes a triangular form in each symmetry sector~\cite{RubioGarciaPRA106}. In the $M=1$ sector, it is convenient to introduce the basis operators
\begin{equation}
O_m \equiv |m\rangle\langle m-1|,
\qquad m=-j+1,\dots,j.
\end{equation}
In this basis, the action of the Liouvillian reads
\begin{equation}
\mathcal L\,O_m
=
\lambda_m\,O_m
+
\eta_m\,O_{m-1},
\end{equation}
where the diagonal elements $\lambda_m$ correspond to the exact eigenvalues in this sector, and $\eta_m$ are nonzero coefficients determined by the dissipative coupling~\cite{RubioGarciaPRA106}. Importantly, the spectrum exhibits exact pairwise degeneracies,
\begin{equation}
\lambda_m = \lambda_{2-m},
\end{equation}
except for the extremal modes, reflecting the presence of second-order exceptional points in the thermodynamic limit.

\paragraph*{Resolvent structure.}

The source operator takes the form
\begin{equation}
X \equiv \rho_0 J^+
=
\sum_{m=-j+1}^{j} s_m O_m,
\qquad
s_m = p_{m-1} a_m,
\end{equation}
with $a_m = \sqrt{(j+m)(j-m+1)}$. Writing
\begin{equation}
(i\omega - \mathcal L)^{-1} X
=
\sum_{m=-j+1}^{j} y_m(\omega)\,O_m,
\end{equation}
the coefficients $y_m(\omega)$ satisfy the recursion relation
\begin{equation}
\Delta_m(\omega)\,y_m(\omega)
-
\eta_{m+1}\,y_{m+1}(\omega)
=
s_m,
\qquad
\Delta_m(\omega) = i\omega - \lambda_m,
\end{equation}
with boundary condition $y_{j+1}=0$. This can be solved iteratively, yielding the exact expression
\begin{equation}\label{eq:ym_expression}
y_m(\omega)
=
\sum_{n=m}^{j}
\frac{
s_n\,\prod_{k=m+1}^{n}\eta_k
}{
\prod_{k=m}^{n}\Delta_k(\omega)
}.
\end{equation}

The emission spectrum is then given by
\begin{equation}
S(\omega;\rho_0)
=
\frac{1}{\pi}\,\mathrm{Re}\sum_{m=-j+1}^{j} a_m\,y_m(\omega).
\end{equation}

{\color{black}
\paragraph*{Second-order pole contributions.}

The structure of Eq.~(A.\ref{eq:ym_expression}) implies that the resolvent contains contributions involving products of denominators $\Delta_k(\omega)$. In particular, for pairs of indices related by the degeneracy condition $\lambda_m=\lambda_{2-m}$, one obtains terms of the form
\begin{equation}
\frac{1}{\bigl(i\omega-\lambda_m\bigr)^2},
\end{equation}
corresponding to genuine second-order poles. These arise whenever both members of a degenerate pair, as well as the coupling between them, are present in the expansion.

For the infinite-temperature state,
\begin{equation}
\rho_0 = \frac{\mathbb I}{2j+1},
\end{equation}
one has $s_m \propto a_m$, so that all coefficients in the expansion are generically nonzero. As a result, the spectrum contains contributions from the full triangular chain, including the second-order pole terms associated with the degenerate pairs.


The presence of second-order poles at $p=1$ reflects the exceptional structure of the Liouvillian in this limit. However, the resulting emission spectrum is generally a superposition of many contributions with different decay rates within the $M=1$ sector. Consequently, it does not reduce to a single simple super-Lorentzian profile, but rather to a more complex rational lineshape centered at $\omega=h$.

This should be contrasted with the steady-state case, where the source operator $\rho_{\mathrm{ss}}J^+$ selects a single extremal mode that remains non-defective, leading to a purely Lorentzian spectrum despite the presence of exceptional points in the Liouvillian. This highlights the crucial role of the observable and the initial state in determining whether exceptional-point physics is visible in spectroscopic quantities.
}

\subsection{Near-degenerate modes and effective second-order poles}
\label{app:near_EP}

In this appendix we clarify how nearly degenerate Liouvillian modes at finite system size give rise to an effective super-Lorentzian contribution in the emission spectrum, even though the Liouvillian remains strictly diagonalizable.

We consider the resolvent form of the spectrum,
\begin{equation}
S(\omega)
=
\frac{1}{\pi}\,
\mathrm{Re}\,
\mathrm{Tr}\!\left[
J_-\, (i\omega-\mathcal L)^{-1}(\rho_{\mathrm{ss}}J_+)
\right],
\end{equation}
and focus on the contribution of a pair of Liouvillian eigenmodes whose eigenvalues approach each other as the system size increases. Denoting $z=i\omega$, we write their contribution as
\begin{equation}
F(z)
=
\frac{R_+}{z-\lambda_+}
+
\frac{R_-}{z-\lambda_-},
\label{eq:F_pair}
\end{equation}
where $\lambda_\pm=\lambda_0\pm\delta$, with $\delta$ small, and $R_\pm$ are the corresponding residues, which depend on the overlap of the source and detector operators with the eigenmodes.

Near an exceptional point, the associated eigenvectors become nearly parallel, and the residues $R_\pm$ exhibit a singular dependence on the splitting $\delta$. In particular, one can write
\begin{equation}
R_\pm
=
\pm \frac{B}{2\delta}
+
\frac{A}{2}
+
O(\delta),
\end{equation}
where $A$ and $B$ are finite coefficients determined by the observable. Substituting into Eq.~\eqref{eq:F_pair}, we obtain
\begin{equation}
F(z)
=
\frac{\frac{A}{2}+\frac{B}{2\delta}}{z-\lambda_0-\delta}
+
\frac{\frac{A}{2}-\frac{B}{2\delta}}{z-\lambda_0+\delta}.
\end{equation}
Expanding for small $\delta$,
\begin{equation}
\frac{1}{z-\lambda_0\mp\delta}
=
\frac{1}{z-\lambda_0}
\pm
\frac{\delta}{(z-\lambda_0)^2}
+
O(\delta^2),
\end{equation}
one finds
\begin{equation}
F(z)
=
\frac{A}{z-\lambda_0}
+
\frac{B}{(z-\lambda_0)^2}
+
O(\delta).
\label{eq:F_expansion}
\end{equation}

Thus, in the limit $\delta\to 0$, the contribution of a nearly degenerate pair of simple poles approaches that of a second-order pole. In the time domain, this corresponds to the well-known relation
\begin{equation}
\frac{e^{(\lambda_0+\delta)t}-e^{(\lambda_0-\delta)t}}{2\delta}
=
t\,e^{\lambda_0 t}
+
O(\delta^2),
\end{equation}
so that a suitable linear combination of two exponentials approaches the Jordan-block form $(a + b t)e^{\lambda_0 t}$.

We emphasize that at any finite system size the spectrum remains a sum of simple poles. However, when the splitting $\delta$ becomes sufficiently small and the residues scale as above, the combined response of the pair is accurately captured by a Lorentzian plus super-Lorentzian form, corresponding to the leading terms in Eq.~\eqref{eq:F_expansion}. This provides a quantitative justification for interpreting the observed non-Lorentzian features as finite-size precursors of exceptional points.

}

\end{appendix}





\bibliography{references_Linblad.bib}


\end{document}